\documentclass[%
reprint,
superscriptaddress,
 amsmath,amssymb,
 aps,
]{revtex4-2}
\usepackage{graphicx}
\usepackage{dcolumn}
\usepackage{bm}
\usepackage[version=4]{mhchem}
\usepackage{chemstyle}
\usepackage{siunitx}
\usepackage{xcolor}
\usepackage{cleveref}
\begin{document}


\title{Microtube self-assembly leads to conformational freezing point depression}

\author{Tatiana Komarova}
\affiliation{Van 't Hoff Laboratory for Physical \& Colloid Chemistry, Utrecht University, Padualaan 8, 3584 CH Utrecht, the Netherlands}
\author{Thomas Zinn}
\altaffiliation[Current addres: Diamond Light Source Ltd, Harwell Science and Innovation Campus, Didcot, Oxfordshire OX11 0DE UK]{ESRF - The European Synchrotron, 38043 Grenoble, France}
\author{Theyencheri Narayanan}
\affiliation{ESRF - The European Synchrotron, 38043 Grenoble, France}
\author{Andrei V. Petukhov}
\affiliation{Van 't Hoff Laboratory for Physical \& Colloid Chemistry, Utrecht University, Padualaan 8, 3584 CH Utrecht, the Netherlands}
\author{Jasper Landman}
\email{Jasper.Landman@wur.nl}
\affiliation{Laboratory for Physics \& Physical Chemistry of Foods, Wageningen University \& Research, Bornse Weilanden 9, 6708 WG Wageningen, the Netherlands}

\date{\today}

\begin{abstract}
Concentric microtubes of $\beta$-cyclodextrin and SDS grow from the outside in and melt from the inside out. In this study we observe the structure of this hierarchical self-assembling system using in situ small angle X-ray scattering, and relate it with enthalpic changes driving the self-assembly process on the molecular level. We find that the conformation of the crystalline bilayer affects the saturation concentration, providing an example of a phenomenon we call conformational freezing point depression. We propose a model based on freezing point depression, well known from undergraduate physics, and use it to explain the energetics of this hierarchical system, and giving access to material properties without free parameters.

\end{abstract}


\maketitle

\section{Introduction}
Self-assembly has long been recognized as a promising strategy for fabricating hierarchical, multiscale materials such as carbon nanotubes \citep{shimoda2002self}, peptide nanotubes \citep{Hamley2014, Valery2011}, mesoscopic peptide-formed fibers \citep{Aggeli2001HierarchicalFibers, o2011multi, whitesides2002self}, compositional tubule networks \citep{Wong2000}. However, the application of self-assembly mechanisms has been limited by the need for precise tuning of the free energy landscape towards the desired function. To address this, carefully designed colloidal matter with DNA-encoded directed bonds \citep{nykypanchuk2008dna}, as well as hydrophobic \citep{chen2011directed} or magnetic \citep{sacanna2012magnetic} patches have been developed. Yet, in a hierarchical material, the free energy is embedded not only in its microscopic, molecular structure, but also in its mesoscopic conformation. As such, it becomes challenging to predict how a system behaves given how the free energy landscape is dependent on that mesoscopic conformation. Here, we show that we can directly relate the mesoscopic structure of a hierarchical material to enthalpic changes occurring at both the molecular and conformational levels. 

\begin{figure}[tbp]
    \centering
    \includegraphics[width=0.9\linewidth]{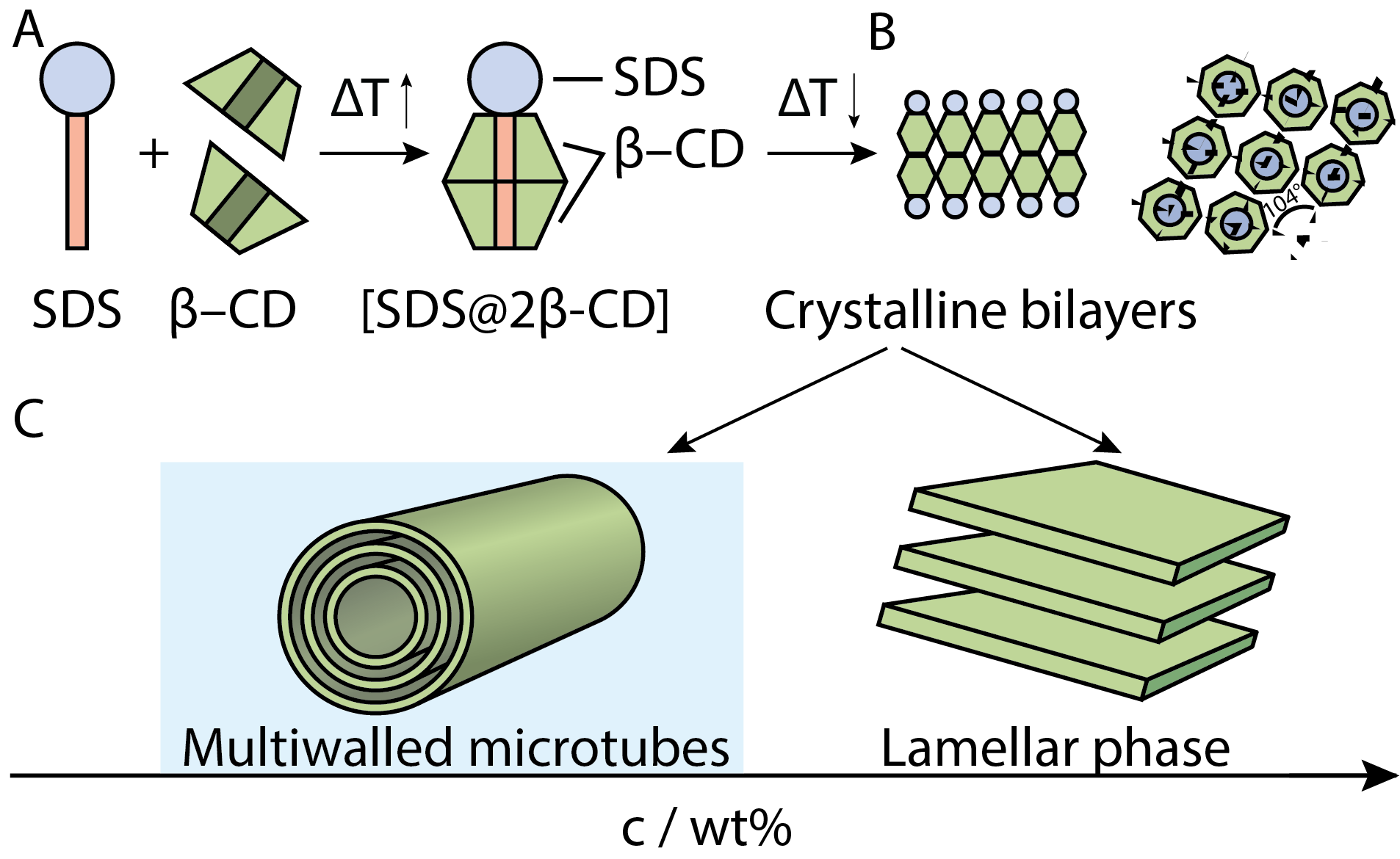}
    \caption{\textbf{Self-organization of $\beta$-CD and SDS into supramolecular aggregates.} (\textbf{A}) The hydrophobic effect drives the formation of \ce{SDS}@2$\beta$-CD complexes \citep{Jiang2011}. (\textbf{B}) Complexes assemble into bilayers, forming a two-dimensional rhombic crystalline membrane via inter-cyclodextrin hydrogen bonding. The rhombic lattice, resulting from the 7-fold symmetry of the $\beta$-CD molecules, maintains a 104$^{\circ}$ angle to maximize inter-cyclodextrin hydrogen bonding, \citep{Yang2017}. (\textbf{C}) Bilayer sheets assemble into superstructures depending on sample concentration \citep{Jiang2011}. Our study focuses on characterizing tube structures (highlighted in blue).}
    \label{fig:schem}
\end{figure}

Recently, the self-assembling system of cyclodextrins (CDs) and sodium dodecyl sulfate (SDS) has demonstrated a remarkable diversity in its mesoscale structures \citep{Jiang2010, Jiang2011, Jiang2011a, Yang2017, Araujo2021}. Depending on the concentrations  of $\beta$-CD and SDS in aqueous solutions, lamellae, multiwalled microtubes as illustrated in \ref{fig:schem}, have been reported to form \citep{Jiang2011}. Along with these phases, \citet{Jiang2011} proposed the existence of rhombic dodecahedra at lower concentrations. \citet{Yang2017} characterized the internal structure of these supramolecular phases, highlighting their similarities to protein- and peptide-based aggregates. The tubular phase is particularly appealing as micro-carriers for controlled drug release \citep{wang2021titanium}, 1D artificial colloid confinement \citep{Jiang2013ab, ouhajji2018wet} and can serve as a model system for fundamental research on hierarchical self-assembly \citep{sun2017nanostructures}.

The self-assembly process of SDS and $\beta$-CD begins with the non-covalent host-guest interactions, yielding pocket-like inclusion complexes with 1:2 stoichiometry denoted by \ce{SDS}@2$\beta$-CD \citep{brocos2010similarities, Jiang2011, valente2014formation}. These complexes subsequently arrange into a rhombic crystalline bilayer membrane with robust unit cell parameters, giving rise to diverse superstructures. Additionally, \ce{SDS}@2$\beta$-CD aqueous solutions are easy to prepare, making them an ideal model system to study the effect of the mesostructural free energy landscape on self-assembly.

Multiple length scales involved in this system require structural analysis techniques that can probe them simultaneously. Synchrotron (ultra)small angle X-ray scattering (SAXS) is ideal for in situ structural analysis of such hierarchical systems. Advanced SAXS instrumentation achieves resolutions from \SI{1}{\nano\meter} to a few microns, enabling the structural elucidation over many orders of magnitude \citep{narayanan2020synchrotron}. As SAXS patterns can provide information on the shape and size of the constituents within the available scattering vector range, this experimental technique is highly suitable for studying hierarchical microtube self-assembly. 
By analyzing SAXS profiles of samples at varying temperatures, we can identify features that enable the determination of the microtube diameter, as well as the inter- and intrabilayer periodicity. Through such analysis, we can gain insight into how the microtube structure on different hierarchical levels is influenced by changes in temperature and concentration.

In this article, we aim to explore the thermodynamic aspects of self-assembly of \ce{SDS}@2$\beta$-CD microtubes. Solutions of varying \ce{SDS}@2$\beta$-CD concentrations were examined at different temperatures using synchrotron SAXS. In our previous study, \citet{Ouhajji2017} probed the structural properties of the system as a function of concentration at a fixed temperature. Furthermore, \citet{Landman2018InwardMembranes} studied the kinetics of microtube formation after a temperature quench. Here, we perform a far wider detailed structural characterization of tubular aggregates, exploring temperature- and concentration-dependent microtube geometrical parameters. The obtained SAXS results enable direct access to nanoscopic energies involved in self-assembly processes, as opposed to the bulk thermodynamic properties \cite{Araujo2022}, revealing valuable insights into the mesoscopic microtube structure and constructing a partial microtube stability diagram.

In particular, we observe correlations between melting temperature, overall microtube diameter and \ce{SDS}@2$\beta$-CD complex concentration. We model this relationship and find that the overal conformation of the complex crystalline bilayer affects the saturation concentration --- providing an example of what can essentially be called conformational freezing point depression.

\section{Results}
\subsection{Characterisation of temperature-controlled SAXS}

\begin{figure*}[tbp]
    \centering
    \includegraphics[width=\linewidth]{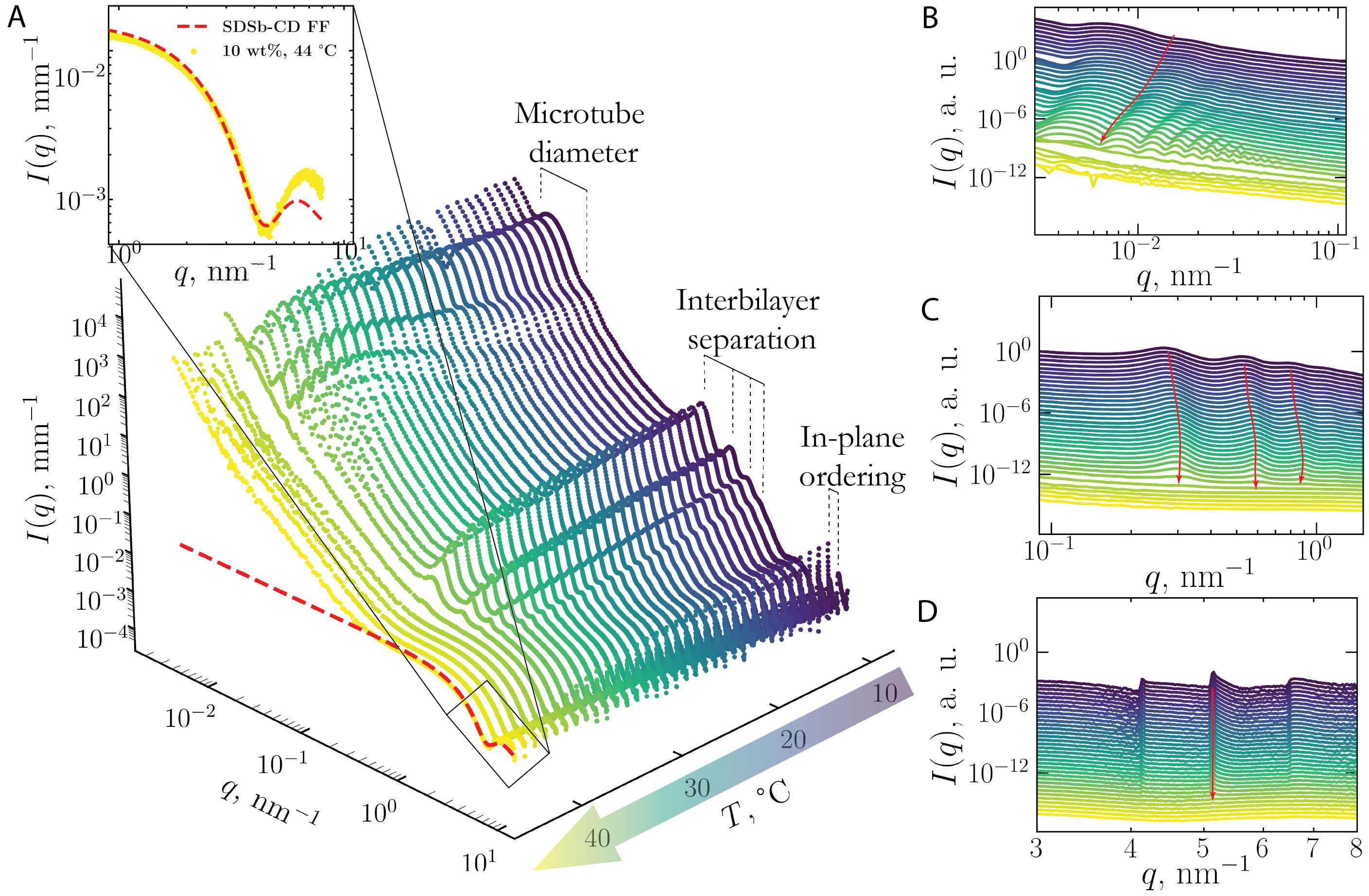}
    \caption{\textbf{Evolution of temperature controlled SAXS} from a \SI{10}{wt\percent} sample. (\textbf{A}) Scattering intensity versus modulus of scattering vector with an arrow indicating the temperature change direction. The dashed red line represents the best fit of the experimental data modeled by the form factor of a \ce{SDS}@2$\beta$-CD complex. Three main levels of the microtube organization are highlighted on the right: (\textbf{B}) oscillations related to the microtube radius at low $q$, (\textbf{C}) intermediate $q$ region reveals regular wall spacing, (\textbf{D}) 2D crystal structure of the bilayer membrane at large $q$ with the peak corresponding to the $[\overline{1} 1]$ crystallographic direction in 2D rhombic lattice. Scattering curves in (\textbf{B})-- (\textbf{D}) are offset, and red arrows indicate specific scattering extremes.}
    \label{fig:temp10wt}
\end{figure*}

Static upramp SAXS experiments were conducted at the TRUSAXS beamline ID02, European Synchrotron Radiation Facility \citep{Narayanan2022} on samples of varying concentrations to investigate the temperature range encompassing microtube melting temperature. Data is available from ESRF\citep{Petoukhov2024Dataset}, and extensive details of the experimental setup and analysis are thoroughly described in the supplementary materials.

\ref{fig:temp10wt} depicts 1D SAXS measurements of a 10 wt\% sample: curves display the sample scattering intensity $I(q)$ as a function of the modulus of scattering vector $q$ for different temperatures $T$. The 3D diagram illustrates temperature-induced sample restructuring across spatial length scales from \SI{2.6E-3}{\per\nano\meter} to \SI{8}{\per\nano\meter}, which nominally corresponds to a real-space range of \SI{0.8}{\nano\meter} to \SI{2.4}{\micro\meter}. 

Focusing on a specific range of $q$ allows for the quantitative analysis of structural changes in the sample. The low $q$ oscillations observed in \ref{fig:temp10wt}B reveal the typical diameter of the microtubes. At low temperatures, these undulations are wide and shallow, indicating the presence of polydisperse and densely packed microtubes. As the temperature increases, higher order oscillations appear, signifying the microtubes becoming more uniform in their size. The shift in the minima to lower $q$ values indicates an increase in the mean microtube diameter. The sudden disappearance of numerous low $q$ oscillations, coupled with a significant decrease of $I(q)$ at higher temperatures, points towards the disintegration of the microtubes into separate complexes. Beyond the melting point, oscillations in this $q$ region are replaced by a power law decay ($\approx q^{-2}$), and the intensity at $q < \SI{0.1}{\per\nano\meter}$ drops dramatically. A similar behavior has been observed in the study by \citet{Landman2018InwardMembranes} and is likely due to a residual amount of flat objects, potentially consisting of fragments of membranes. At even higher temperatures, the recorded scattering patterns remain unchanged without many of scattering features. These scattering curves do not have any specific fingerprints except for a minimum at $\approx$ \SI{4.5}{\per\nano\meter}. 

At larger $q$ values, the scattering curve can be fitted with a form factor of a \ce{SDS}@2$\beta$-CD complex. The dashed red line in Figure \ref{fig:temp10wt} shows the best fit result for the experimental data taken at \qty{44}{\celsius}, which was obtained using molecular dynamics simulations that have been thoroughly described in previous studies \citep{chen2014validating, knight2015waxsis}.

Descending to the lower level of the microtube hierarchical self-assembly depicted in \ref{fig:temp10wt}C, it is observed that pseudo-Bragg peaks are present in the microtube scattering patterns. These peaks refer to the specific distance $d$ between walls inside a multiwalled tube that can be expressed by Bragg equation:
\begin{equation}
    2 \langle d \rangle \textrm{sin} \theta = n \lambda,
    \label{eq:bragg}
\end{equation}
where $2\theta$ is the scattering angle, $\langle d \rangle$ - interbilayer separation between microtube walls, and $n$ is an integer.

Higher harmonics with $n>1$ of the repetitive separation of the microtube layers can be observed at higher $q$ values, as shown in \ref{fig:temp10wt}C. With an increase in temperature, the lamellar peaks shift insignificantly to higher $q$ values, indicating wall convergence. As the transition temperature is approached, the peak position shifts to lower $q$ values and its intensity gradually decreases. According to \citet{Landman2018InwardMembranes}, the inner layers of the multilayered tubes are less energetically stable due to the higher bending energy of the crystalline membrane they consist of, which leads to their melting at lower temperatures compared to the outer layers. This results in a drop in the lamellar peak intensity and, ultimately, its disappearance. The melting of the microtubes can also be observed in polarized light optical microscopy images taken from a sample at different temperatures, as shown in the supplementary materials. The gradual disintegration of the inner cylinders is accompanied by the release of pocket-like SDS@2$\beta-$CD complexes into the solution. 


The highest $q$ values in \ref{fig:temp10wt}D provide insight into the internal crystalline structure of the microtubes. Previous studies, such as the work of \citet{Yang2017}, have indicated that the microtubes are comprised of a crystalline bilayer membrane in which \ce{SDS}@2$\beta$-CD complexes are arranged in a 2D rhombic lattice that optimises the positioning of hydrogen bonds between the seven-fold symmetric cyclodextrins. The temperature-dependent behavior of three tooth-shaped peaks, corresponding to the $[1 0]$, $[\overline{1} 1]$, and $[1 1]$ in-plane crystallographic directions, is presented in \ref{fig:temp10wt}D. The peak positions and, hence, rhombic unit cell parameters remain constant throughout the heating process. However, the peak intensities gradually decrease with increasing temperature and ultimately disappear around the melting temperature, indicating the absence of crystalline material in the observed volume.

The SAXS profiles obtained in this study enable both qualitative tracking of temperature-induced sample evolution and quantitative analysis of changes in various structural features of the microtubes, such as the outermost radius, number of walls, and inter-wall distance, as a function of temperature.

\subsection{Temperature-concentration phase space}
The SAXS patterns for all concentrations show the same structural transformations: initially, polydisperse, densely populated multilayered tubes transform into more uniform objects, continuously breaking up into \ce{SDS}@2$\beta$-CD single complexes. The absolute value of the transition point is dependent on the sample concentration. The transition temperature for each sample was determined using the Porod invariant within a specific $q$ region, and the procedure is described in detail in the supplementary materials. 

The existence of various phases in aqueous solutions of \ce{SDS}@2$\beta$-CD complexes at room temperature has been reported previously in the study of \citet{Yang2017}. In a more recent study, the phase behavior of pH-responsive surfactant-cyclodextrin complex mixtures was reported by \citet{araujo2022hierarchical}. 

\begin{figure}[tbp]
    \centering
    \includegraphics[width=1\linewidth]{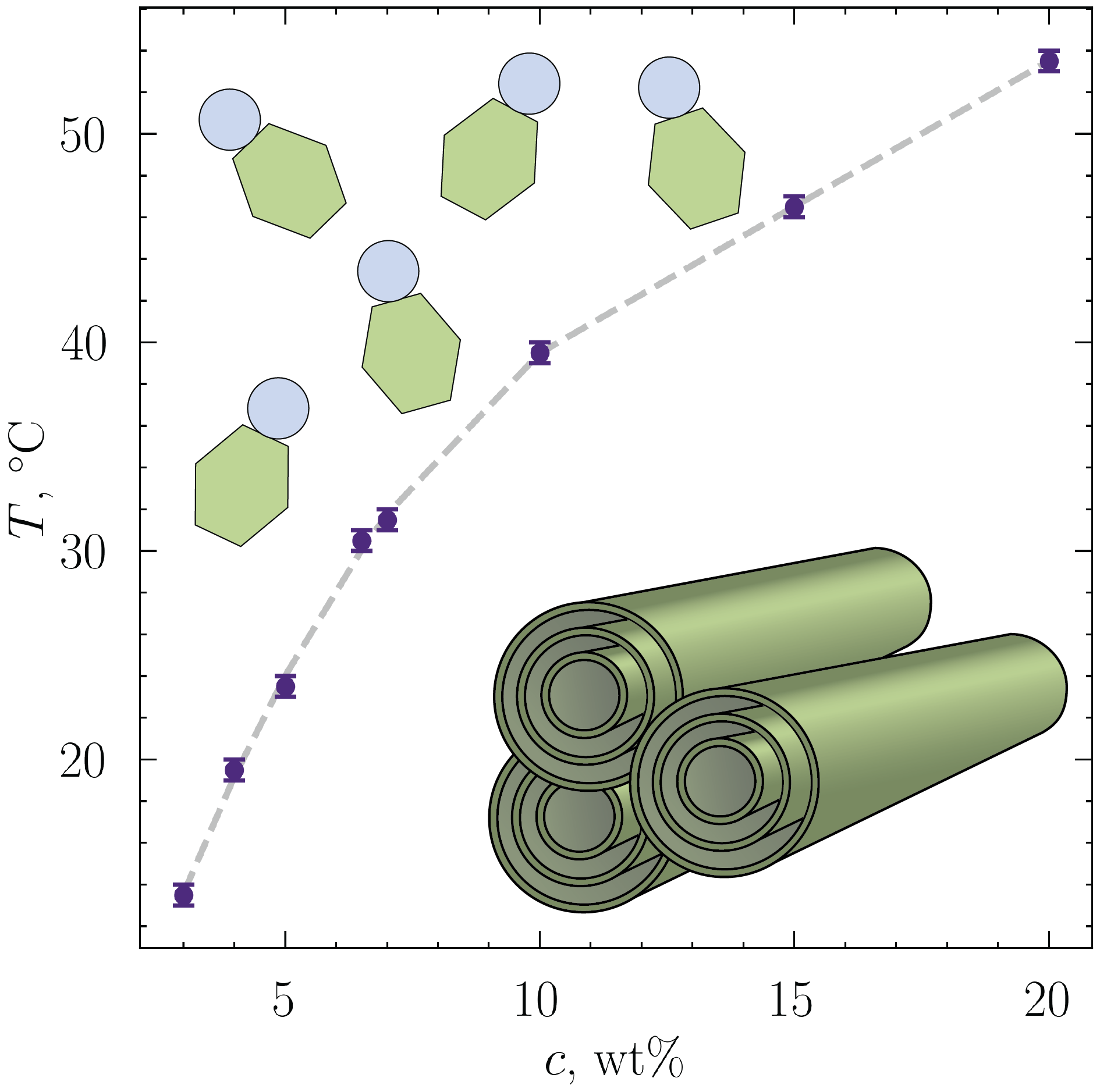}
    \caption{\textbf{Microtube melting point versus sample concentration.} Circles represent transition points between microtubes and \ce{SDS}@2$\beta$-CD complexes at various sample concentrations. The grey dashed line denotes the binodal.}
    \label{fig:Tmelt}
\end{figure}

By analyzing the upramp SAXS series, we have reconstructed a partial thermodynamic diagram in temperature-concentration coordinates, as shown in \ref{fig:Tmelt}. At temperatures exceeding melting temperatures, no tubular aggregates were found, we exclusively observed \ce{SDS}@2$\beta$-CD complexes. To the best of our knowledge, it is the first time that the full temperature-dependent behavior of our system has been demonstrated. 

In general, an increase in sample concentration leads to a higher transition temperature. As previously observed in the temperature series of the 10 wt\% sample (\ref{fig:temp10wt}), only the form factor of single complexes is detected at temperatures exceeding the melting point. The same scattering curve evolution was observed for all concentrations tested. Therefore, the dashed line in \ref{fig:Tmelt} separates the region of tubular phase existence and the region where only free single \ce{SDS}@2$\beta$-CD complexes are present. 

The transition point curve in \ref{fig:Tmelt} is located near the physiological temperature. The reversible transition between tubes and single complexes around the temperature of the human body could be exceptionally advantageous for controlled drug/dye release. Due to the biocompatibility of $\beta$-cyclodextrin, an increasing number of studies, such as those by \citet{datz2018biocompatible} and \citet{abbas2022galangin}, employ this cyclic oligosaccharide as a basic component to create nanocarriers and drug delivery systems.

\subsection{Structural parameters}

\begin{figure*}[p]
    \centering
    \includegraphics[width=\linewidth]{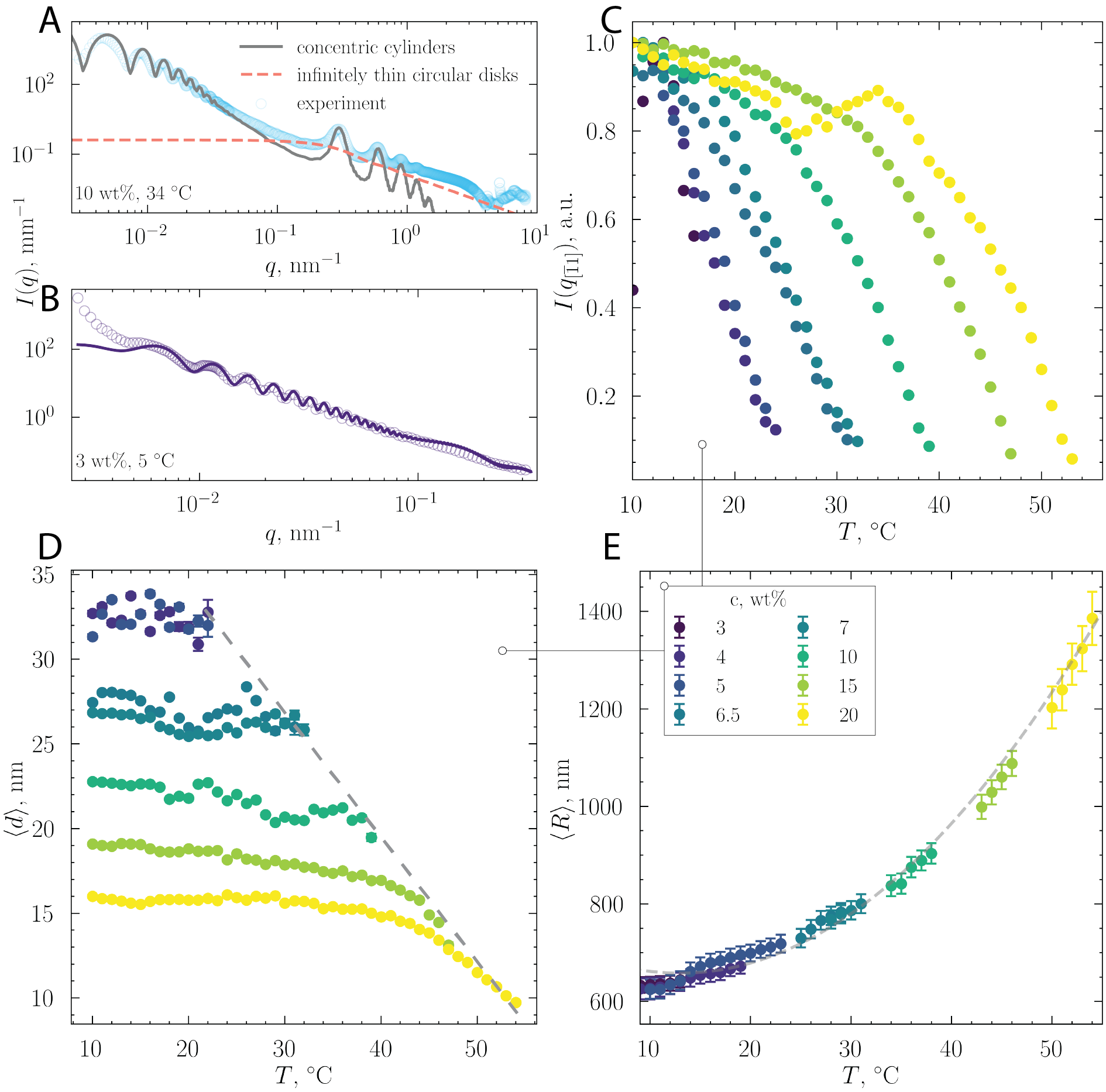}
    \caption{\textbf{Tracing the microtube parameters.}  (\textbf{A}) Azimuthally averaged scattering curve of the \qty{10}{wt\percent} sample taken at \qty{34}{\celsius} (blue) and the linear combination of scattering functions of concentric hollow cylinders (grey) and disks (salmon). (\textbf{B}) Experimental data taken from a \qty{3}{wt\percent} sample recorded at \qty{5}{\celsius} (circles)  and the best fit obtained using the linear combination of concentric cylinders and disks (line). (\textbf{C}) Scattering intensity of the crystalline membrane $[\overline{1}1]$ peak as a function of the sample temperature. (\textbf{D}) Interbilayer separation at different temperatures for various concentrations. For low concentrations, its value does not depend on the temperature and fluctuates around the average value. For higher concentrations, a certain decaying trend with the temperature increase is observed. (\textbf{E}) Microtube outermost radius for various concentrations. The values found from the fitting procedure follow a uniform temperature dependence. The grey dashed line corresponds to the parabolic fit line with empirical parameters of the best fit result.} 
    \label{fig:params}
\end{figure*}

The scattering data were analysed using a structural model of long concentric cylinders \citep{Livsey1987, Paineau2016, Ouhajji2017} and coexisting thin circular disks \citep{kratky1949diffuse}. The model is described in detail in the supplementary materials. In \ref{fig:params}A,B we show how the model can fit a scattering pattern at \qty{10}{wt\percent} and \qty{3}{wt\percent} respectively. From the model we can quantitatively describe the scattering contributions of the multiwalled tubules, up to a scattering vector of $q=\qty{1.8}{\per\nano\meter}$. At higher $q$ the sample scattering is governed by the crystalline organisation of complexes within a bilayer, which is not taken into account in the model. 

\ref{fig:params}B demonstrates the ability of the proposed model to describe a SAXS curve taken from a 3 wt\% sample at \SI{5}{\celsius}. \citet{Jiang2011} found a polyhedral capsid phase at $\leq 6$ wt\% in cryo-EM micrographs. We find no evidence for this phase in our experiments. SAXS patterns of samples recorded in this concentration regime all show the existence of the microtubular phase, as shown in the supplementary materials.

In \ref{fig:params}C, we plotted the relative intensity of the most prominent intra-bilayer peak (in the $[\overline{1} 1]$ crystallographic direction) against temperature. Assuming that the tubular phase is the primary source of crystalline material in the system, this trend allows us to estimate the fraction of complexes that are released into the solution. For all concentrations, the intensity decay is the same, which suggests that the energetics of microtube disintegration is concentration-independent.

The distance between microtube walls can also be traced in the fitting procedure. As shown in \ref{fig:params}D, the interbilayer separation is relatively constant for low concentrations within the available temperature range, whereas for higher concentrations, its value decreases with increasing temperature. In our previous work \citep{Landman2018InwardMembranes}, we proposed an expression for this quantity

\begin{equation}
    d = \sqrt{\frac{k_BT\sigma^2}{\rho_s}\frac{4\pi R^2}{\sqrt{3}a_0 \kappa c}}.
    \label{eq:dist}
\end{equation}
Here, $k_B$ is the Boltzmann constant, $\sigma$ represents the surface charge number density of the membrane, $\rho_s$ is the salt number density, $a_0$ is the interfacial area occupied by a single \ce{SDS}@2$\beta$-CD complex, and $c$ is the number density of \ce{SDS}@2$\beta$-CD complexes.

The temperature term $T$ in the numerator in \cref{eq:dist} intuitively suggests a positive correlation in the average distance with increasing temperature. However, highly concentrated samples demonstrate the opposite trend. The temperature driven decrease of $d$ can be explained as an effect of the ionic strength increased by free \ce{SDS}@2$\beta$-CD complexes. As the temperature increases, the inner layers of the microtubes disintegrate, providing additional macroions to the salt reservoir, screening the remaining electric double layer repulsions between the layers that remain.

\ref{fig:params}E represents the outermost radius values obtained from the fitting procedure. Resulting values are temperature-dependent and exhibit a uniform master curve that is independent of the sample concentration. A parabolic fit is shown as the grey dashed line, indicating that the microtube size increases with increasing temperature. We can infer from \ref{fig:params}E that the microtube size is determined by the competition between energy contributions that define the self-assembly path, which are independent of the sample concentration and change only with temperature.

\subsection{Microtube melting model}
In our earlier work\citep{Landman2018InwardMembranes} we found that the mechanism is driven by the gain in free energy achieved by a crystalline bilayer of SDS@$\beta$-CD complexes when it closes to form a cylinder. This happens at a definite size determined by optimizing the free energy gain per unit interface and the bending penalty. The outcome is a nucleation-dominated inward growth until a space-filling structure is formed.

In current temperature ramp experiments, we found that the innermost cylinders have the lowest melting temperatures. Essentially, when slowly increasing the temperature of a microtube system, the tubes melt from the inside out. This aligns with the idea that tightly curved inner cylinders gain less free energy upon closure due to increased bending energy.

Two key physical parameters influencing melting behaviour are the the bond enthalpy of a complex being incorporated into the bilayer, and the bending modulus of that bilayer. Although direct access to these parameters is unavailable, we monitored key observables in the SAXS analysis of microtube melting: macroscopic complex concentration, melting temperature of the outermost cylinder, and outermost cylinder radius. 

The central assumption in the model proposed here is that the bond enthalpy of a complex being incorporated into a bilayer is temperature-dependent, but the bending modulus, at least to the first order, is constant. The crystalline nature of the membrane forming the tubular aggregates distinguishes it from soft matter composed of lipid bilayer membranes, as the latter is easily deformed, having a bending modulus typically not exceeding a few tens of $k_{\text{B}}T$ \citep{Nagle2017, Eid2020}. While even minor temperature variations can significantly impact bilayer membranes, the bending modulus of \ce{SDS}@$\beta$-CD microtubes is presumably notably higher and can be reasonably assumed to remain constant, irrespective of temperature changes. According to our previous study \citep{Landman2018InwardMembranes}, the enthalpic contribution is expected to be significantly smaller, typically on the order of $k_{\textrm{B}}T$, as it characterizes the energy associated with the bonding interactions between cyclodextrin molecules \citep{Wang2022, Bhat2023}.

The (Helmholtz) free energy of bending per unit interface of a sheet of material bent uniformly along one principal axis is given by
\begin{equation}
    f_\text{bend} = \frac{\kappa}{2}\frac{1}{r^2},
 \end{equation}
where $\kappa$ is the elastic bending modulus of the bilayer \citep{Helfrich1984} and $r$ is the radius of the curvature. Here we have omitted the Gaussian curvature term, which is zero for all flat and cylindrical objects. Upon deforming a bilayer of width $2\pi r$ and length $\ell$ (assuming an approximately rectangular geometry) and closing the cylinder, the line tension along the length of the bilayer is removed. Per unit interface, the free energy gain of this process is given by
\begin{equation}
    f_\text{bond} = -\frac{\tau}{2\pi}\frac{1}{r},
\end{equation}
where $\tau$ is the line tension, i.e. the free energy per unit length that arises from the unpaired bonds at the edge of the cylinder. In terms of microscopic quantities, the line tension $\tau$ scales with the typical bond energy through a lengthscale $\ell_0$ that is on the order of the lattice parameter of the bilayer.

Combining the free energy of bending per unit interface  and the free energy gain of closing the cylinder, and then setting the derivative with respect to $r$ to $0$, we find an optimum cylinder radius given by
\begin{equation}
    r_0=\frac{2\pi \kappa}{\tau}.\label{eq:prefRadius}
\end{equation}

The optimal cylinder radius can be determined experimentally by analyzing SAXS patterns near the melting temperature. \Cref{eq:prefRadius} relates $\kappa$ and $\tau$, connecting mesoscopic and molecular scales.

We now assume that at the melting point, there is an association equilibrium between a \ce{SDS}@2$\beta$-CD complex incorporated into a tube, and a \ce{SDS}@2$\beta$-CD complex floating free in solution. This equilibrium is reminiscent of a surfactant solution that can self-assemble into micelles. As such, we have the equilibrium condition
\begin{equation}
    \mu_\text{A}^\text{(aq)} = \mu_\text{A}^\text{(c)},
\end{equation}
with the superscripts now denoting the aggregation state of the complex. Assuming the ideal behaviour of the solution, the chemical potential can be written as
\begin{equation}
 \mu_\text{A}^\text{(aq)} = \mu_\text{A}^{\standardstate\text{(aq)}} + k_\text{B}T \log \frac{x_\text{A}}{x_\text{A}^\standardstate},
\end{equation}
where we have chosen an arbitrary concentration to act as the reference state for which $\mu_\text{A}^\standardstate$ holds. The chemical potential of a complex inside the cylinder $\mu_\text{A}^\text{(c)}$ can be seen as the reference chemical potential of a pure component, $\mu_\text{A}^\star$. We can identify $\Delta_{\text{r}}g_{\text{A}}\equiv\mu_{\text{A}}^{\star}-\mu_{\text{A}}^{\standardstate\text{(aq)}}$ as the difference in reference chemical potentials as the standard molar Gibbs free energy of the reaction $\ce{A(aq) <=> A(c)}$, which in turn can be split into an enthalpic and an entropic contribution.

\begin{equation}
    \log\frac{x_\text{A}}{x_\text{A}^\standardstate} = -\frac{\Delta_\text{r}g_\text{a}}{k_\text{B}T} = \frac{\Delta_\text{r}h_\text{A}(T)}{k_\text{B}T} - \frac{\Delta_\text{r}s_\text{A}}{k_\text{B}}.\label{eq:standardGibbs}
\end{equation}
At this point, we take \cref{eq:standardGibbs} evaluated at the reference concentration $x_\text{A}^\standardstate$,
\begin{equation}
 \log\frac{x_\text{A}^\standardstate}{x_\text{A}^\standardstate} = 0 = \frac{\Delta_\text{r}h_\text{A}(T)}{k_\text{B}T^\standardstate} - \frac{\Delta_\text{r}s_\text{A}}{k_\text{B}},
\end{equation}
and subtract it from \cref{eq:standardGibbs} to cancel out the temperature-independent entropy terms. We are then left with 

\begin{equation}
   \log{\frac{x_\text{A}}{x_\text{A}^\standardstate}} = \frac{\Delta_\text{r}h_\text{A}(T)}{k_\text{B}T}-\frac{\Delta_\text{r}h_\text{A}(T^\standardstate)}{k_\text{B}T^\standardstate}.
   \label{eq:freezingpointChange}
\end{equation}

\begin{figure*}[tbp]
    \centering
    \includegraphics[width=0.8\linewidth]{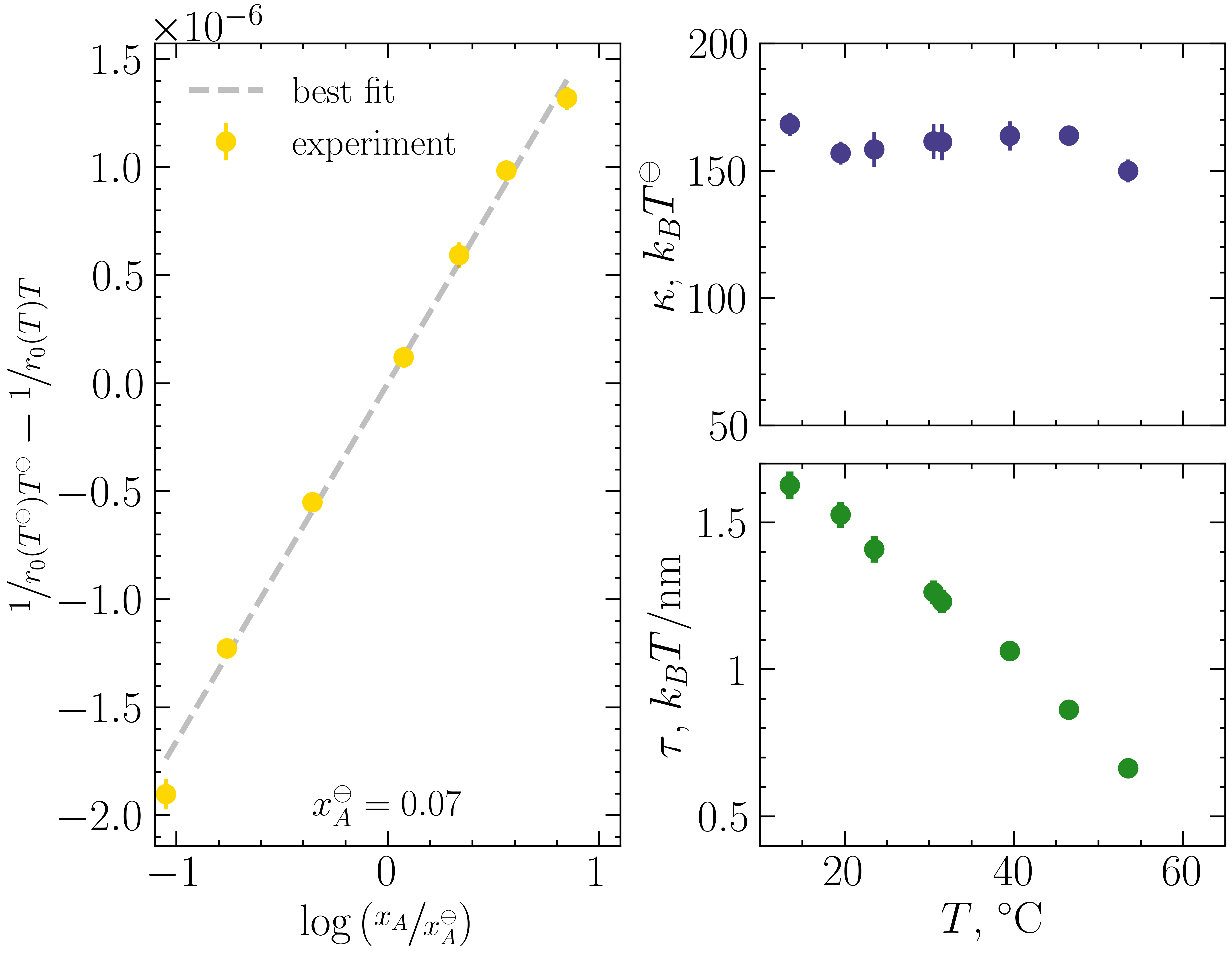}
    \caption{\textbf{Microtubule melting model applied to SAXS experimental data.} (\textbf{A}) Experimental points and a linear trend following the developed model, \cref{eq:eq31}. A 7 wt\% sample was used as a reference ($x_\text{A}^\standardstate$, $T^\standardstate$). (\textbf{B}) Bending modulus in $k_{\textrm{B}}T$ units for differently chosen reference temperature $T^\standardstate$ and (\textbf{C}) line tension values obtained from the model ($T^\standardstate=298$K).}
    \label{fig:kappa}
\end{figure*}

Assuming that the enthalpic contribution $\Delta_\text{r}h_\text{A}$ in \cref{eq:freezingpointChange} is equivalent to the bond energy gained from the complex incorporation into the bilayer, we say that
\begin{equation}
    \Delta_\text{r}h_\text{A}(T) = -\frac{\ell_0 \tau(T)}{2},
\end{equation}
where the factor $1/2$ is introduced because the structure is a bilayer. This can be inserted into \cref{eq:freezingpointChange} to yield
\begin{equation}
    \log{\frac{x_\text{A}}{x_\text{A}^\standardstate}} = \frac{\ell_0 \tau(T^\standardstate)}{2k_\text{B}T^\standardstate} - \frac{\ell_0 \tau(T)}{2k_\text{B}T}.
\end{equation}

While we do not have access to $\tau$ directly, we use the established relationship between the line tension $\tau$ and the experimentally accessible preferential radius $R$, where the bending modulus $\kappa$ is considered constant across the entire temperature range. 
Additionally, the role of line tension in nucleation rate, as discovered in \citep{Landman2018InwardMembranes}, provides an independent $\kappa$ value, offering a parameter-free closure. From the nucleation rate at room temperature, a line tension is about \SI{0.8}{kT\per\nano\meter}.

We insert the relationship given by \cref{eq:prefRadius} into the model, yielding
\begin{equation}
    \log{\frac{x_\text{A}}{x_\text{A}^\standardstate}} = \frac{\ell_0 \pi\kappa }{r_0(T^\standardstate)k_\text{B}T^\standardstate} - \frac{\ell_0 \pi\kappa }{r_0(T)k_\text{B}T}.
       \label{eq:eq31}
\end{equation}

We tested the derived expression against the experimentally determined outer radii as discussed above, where the melting temperature, the outermost radius value, and the sample concentration were considered the three main parameters necessary to access the energetics underlying tube formation. When plotting $[r_0 (T^\standardstate)T^\standardstate]^{-1} - [r_0 (T)T]^{-1}$ as a function of $\log{x_{\ce{A}}/x_{\ce{A}}^\standardstate}$, we obtain a straight line with slope $\ell_0 \pi \kappa$, as can be seen in \ref{fig:kappa}A for a chosen reference concentration of \SI{7}{wt\percent}. The bending modulus can then be isolated directly from the proportionality constant. We verified the assumption of constant $\kappa$ by choosing different sample concentrations as the reference state and obtained the corresponding slope and concurrent bending modulus. 
In \ref{fig:kappa}B we show the fitted bending modulus as a function of the melting temperature of the chosen reference state. Indeed, we see that $\kappa$ is independent of the selected reference state, validating our a posteriori assumption of a constant bending modulus. The ability to get this physical quantity from the SAXS experiment is extremely valuable: usually, more sophisticated methods are implemented, for example \citep{dimova2014recent}.

\ref{fig:kappa}C illustrates the line tension $\tau$ resulting from the fitting procedure. The model assumes that the line tension is essentially the enthalpy contribution: its value should lie around $\approx 1 k_{\textrm{B}}T \textrm{nm}^{-1}$\citep{Landman2018InwardMembranes}, which is in line with current results. Moreover, we observe a decrease in $\tau$ with increasing temperature, consistent with earlier observations about the temperature dependence of hydrogen bonds which are the primary contribution to the line tension in this system \citep{ross1996thermodynamics} -- as temperature rises, intensifying thermal fluctuations result from the decrease in $\tau$, ultimately causing hydrogen bonds between $\beta$-cyclodextrins to break, and microtubes to disassemble. 

As has been already mentioned, the \ce{SDS}@2$\beta$-CD system shares some similarities with natural objects such as proteins and peptides \citep{tanford1980hydrophobic, discher2002polymer, nikolic2009micelle}. The microtube membrane is rigid and crystalline, which is a significant difference from soft matter structures based on bilayers that are easily distorted and have a bending modulus that does not exceed tens of $k_{\textrm{B}}T$. Therefore, while slight temperature deviations can substantially affect soft matter structures based on bilayers, this is not the case for \ce{SDS}@2$\beta$-CD tubes, whose bending modulus value is much larger.

The microtube melting model presented above, despite its simplicity and underlying assumptions, provides valuable insights into the energetic contributions crucial for microtube formation, directly derived from the SAXS experiment. The obtained micro- and macroscopic parameters enable us to manipulate the tube formation and vary the energetic contributions of different nature, providing guidance for designing and optimizing tube formation processes. Importantly, the model can be extended to other systems that operate at different length scales, providing a powerful tool for studying the disassembly and related energetic processes of such systems. For instance, similar models can be applied to SAXS data on peptide nanotubes or other amphiphilic systems forming nanotubes to track the conditions and geometrical parameters of their disassembly and relate them to the energy governing this process.

\section{Conclusion}
In this study, we have investigated the structural response of multiwalled tubular supramolecular aggregates formed in aqueous solutions of SDS and $\beta$-CD as a function of temperature and concentration by means of the SAXS technique. Modeling of SAXS profiles enabled to obtain outer radii, interbilayer separation, and 2D rhombic lattice parameters. 
Temperature-induced microtube assembly/disassembly follows the microtube inward growth proposed by \citet{Landman2018InwardMembranes}.

The outermost radius of the tubes is highly sensitive to the temperature, while \ce{SDS}@2$\beta$-CD complex concentration insignificantly affects this quantity. As temperature increases towards their melting point, the number of walls inside microtubes decreases, and the tube swells. Temperature-dependent outer radius values follow the master curve and scale with the sample concentration. 

We demonstrated an approach to access enthalpic changes driving hierarchical self-assembly directly from structural parameters of supramolecular assemblies. Assuming that microtube formation results from the interplay between bending energy and bond formation, the temperature dependence of the outermost radius sheds light on the energetics of the self-assembly, allowing us to estimate the energies of H-bonds involved in this process. Using the macroscopic quantities (the microtube radius, transition temperature) obtained from the SAXS experiments and the developed model, we derived nanoscopic quantities such as anisotropic bending modulus and line tension. The proposed model describing microtube melting behaviour is comprehensive and can be applied to other systems in which tubular objects are formed as a result of the high energy of hanging edge bonds and bending stiffness of the membrane.

\begin{acknowledgments}
Bonny Kuipers is gratefully acknowledged for his help with the polarization microscopy setup. We acknowledge the European Synchrotron Radation Facility (ESRF) for provision of synchrotron radiation facilities, and the use of beamline ID02 for experiments SC-4987 and SC-5177, and for hosting TK for a traineeship.  
\end{acknowledgments}

\end{document}


\title{Microtube self-assembly leads to conformational freezing point depression \\
Supplementary Materials}

\author{Tatiana Komarova}
\affiliation{Van 't Hoff Laboratory for Physical \& Colloid Chemistry, Utrecht University, Padualaan 8, 3584 CH Utrecht, the Netherlands}
\author{Thomas Zinn}
\altaffiliation[Current addres: Diamond Light Source Ltd, Harwell Science and Innovation Campus, Didcot, Oxfordshire OX11 0DE UK]{ESRF - The European Synchrotron, 38043 Grenoble, France}
\author{Theyencheri Narayanan}
\affiliation{ESRF - The European Synchrotron, 38043 Grenoble, France}
\author{Andrei V. Petukhov}
\affiliation{Van 't Hoff Laboratory for Physical \& Colloid Chemistry, Utrecht University, Padualaan 8, 3584 CH Utrecht, the Netherlands}
\author{Jasper Landman}
\email{Jasper.Landman@wur.nl}
\affiliation{Laboratory for Physics \& Physical Chemistry of Foods, Wageningen University \& Research, Bornse Weilanden 9, 6708 WG Wageningen, the Netherlands}

\date{\today}
\maketitle

\tableofcontents

\section{Materials and methods}
\subsection{Preparation of [SDS@2$\beta$-CD] microtube suspensions}
$\beta$-CD (Sigma-Aldrich, 97\%), SDS (Sigma-Aldrich, $>$99\%) and Milli-Q water were weighed and mixed together in the desired amounts with a constant SDS to the $\beta$-CD 1:2 molar ratio. SDS was used to prepare samples as received. $\beta$-CD was dried to get rid of water as it is known for its hygroscopicity \citep{bilal1995beta}. According to gravimetric analysis of the $\beta$-CD powder before/after dehydration (drying in an oven at 120$^{\circ}C$ for 24 hours), there were 11 $\textrm{H}_2\textrm{O}$ molecules per one sugar molecule. A sample series with \ce{SDS}@2$\beta$-CD concentrations ranging from 3 weight \% (wt\%) to 20 wt\% was prepared. Mixtures were stirred and heated up to 70$^\circ$C, until a turbid solution has changed to a transparent one corresponding to the \ce{SDS}@2$\beta$-CD complex emergence. Then, hot solutions were kept for 48 hours at room temperature for tubular phase formation yielding turbid, viscous gels. The dynamics of the turbidity appearance depends on the concentration: the higher the concentration, the faster the sample becomes turbid and highly viscous. For samples with extremely high concentrations (30, 40 wt\%), the turbidity is less pronounced, which supposedly results from the formations of another, lamellar phase (\citet{Jiang2013a}).

\subsection{Small Angle X-ray scattering}
All scattering experiments have been performed at the Time-Resolved (Ultra-)Small Angle X-ray Scattering (TRUSAXS) beamline ID02, European Synchrotron Radiation Facility \citet{Narayanan2022}. The incident X-ray wavelength was 1.01 \AA. Scattering data were recorded using an Eiger2 4M (Dectris AG) hybrid pixel-array detector at two different sample-to-detector distance: 31 m for ultra-small angles ($q_{\rm min}$ = 0.002 nm$^{-1}$) and 1 m for larger angles ($q_{\rm max}$ = 8 nm$^{-1}$). These distances allow one to cover nominal sizes from 0.8 nm to 2.4 $\mu$m. Measured 2D SAXS patterns were normalized, regrouped and azimuthally averaged to obtain the 1D SAXS profiles as described elsewhere \citep{Narayanan2022}. The background scattering curve was recorded at the same conditions as samples from a capillary with filled Milli-Q water.

Since it is impossible to cover length scales featuring both radius oscillations and crystalline structure in one scattering experiment, the sample-to-detector distance has to be changed after the experiment is finished. Then, the scattering experiment is repeated under the same conditions with a sample from the same vial, which was not modified during the first experiment. That provides identical conditions for all taken scattering curves. Then, the recorded curves are merged to span all nominal length scales of interest. 

\subsection{Heating stage coupled with SAXS}
Samples with different concentrations (in the range from 3 wt\% to 20 wt\%) were loaded into thin walled quartz capillaries with a diameter of 2 mm. Samples were placed in a temperature controlled stage for multiple capillaries.  Capillaries were sealed to prevent solvent evaporation. Since the studied samples are viscous, the shear effect resulting from capillary filling affects the scattering patterns. 2D scattering images are slightly oriented, which makes subsequent data reduction somewhat troublesome. To avoid this, all sealed capillaries were heated to a temperature at which all structural features corresponding to the tubular phase disappear (60$^{\circ}$C) before the experiments. This procedure ensured that the samples were brought to the same molten state. Then, solutions consisting of single complexes were gradually cooled down with a cooling rate of 0.2$^{\circ}$C/min and left for ten minutes at a set temperature to achieve equilibrium before recording the SAXS pattern. A downramp series was recorded in the temperature range of 70$^{\circ}$C-5$^{\circ}$C with a step of 1$^{\circ}$C. The 3 wt\% sample had to be cooled below 2$^{\circ}$C and 30 wt\% required heating above 70$^{\circ}$C. Afterward, a temperature upramp series was recorded covering 5$^{\circ}$C-70$^{\circ}$C temperature range. 
\subsection{Polarized light microscopy}
To study the melting behaviour of tubular aggregates, the prepared suspension was loaded into a rectangular glass capillary (Vitrocom, 0.1$\times$2$\times$50 mm) and sealed by ultraviolet-curing epoxy glue. Nikon Eclipse E400 POL polarising microscope equipped with Linkam heating stage was employed to conduct experiments in the polarization mode to track the presence of birefringence in the sample at various temperatures.

\subsection{Analysis}
Results of the SAXS experiment were visualized and analysed using the software SAXSutilities2 \citep{sztucki_michael_2021} in combination with custom Python scripts.

\subsection{Temperature-dependent SAXS}
Initially, all the samples were heated until their SAXS profiles only displayed scattering features corresponding to individual \ce{SDS}@2$\beta$-CD complexes using a temperature controlled multi-capillary stage. This procedure eliminated any potential shear-induced orientation of the samples that may have resulted from capillary filling. Then, the temperature was subsequently decreased stepwise with \SI{1}{\degreeCelsius} increments and SAXS patterns were recorded at each new temperature with equilibration time of 10 minutes. This process was repeated until the set lowest temperature was reached forming the downramp SAXS series. Following this, the upramp series was recorded in which the sample was initially at a low temperature and then heated until the SAXS patterns remained unchanged.

\section{Modeling scattering pattern}
In addition to the qualitative description presented in Fig. 2, quantitative analysis can be performed by modelling the recorded scattering patterns. The structure factor of interactions is inaccessible due to the combination of repulsive structure factor and hollow morphology resulting in the sharp drop of the SAXS intensity at low $q$ \citep{Glatter1982}. This, along with the analysis performed beyond the first minima of the concentric cylinder form factor, enables the total recorded scattering to be modeled solely based on the form factors of the scattering objects present in the solution. 

SAXS patterns acquired from different concentrations and temperatures below melting temperatures were quantitatively analyzed by fitting to the concentric cylinder model \citep{Livsey1987, Paineau2016, Landman2018InwardMembranes}. The orientationally averaged scattering function of long concentric tubes equidistantly arranged inside each other can be expressed as presented in \cref{eq:ff_tubes}, which has been previously described by \citet{Livsey1987} and \citet{Paineau2016}:

\begin{widetext}
\begin{equation}\label{eq:ff_tubes}
I_{\textrm{tubes}}(q) = n \frac{2 \pi^3 \left< \Delta \rho \right> ^2 L}{q} \times \sum_{m=0}^{N-1} \left(\frac{RJ_1 \left( qR\right)}{q} - \frac{\left(R - md - t_{\text{b}}\right)J_1\left(q \left(R - md - t_{\text{b}}\right)\right)}{q}\right)^2.
\end{equation}
\end{widetext}

Here, $n$ is the volumetric number density of scattering centers, $J_1$ denotes the 1$^{st}$-order Bessel function of the first kind, $\Delta \rho$ is the difference between scattering length densities of cylindrical shells and the solvent (the calculation of the $\Delta \rho$ used to model the experimental data can be found below), $R$ is the outermost radius of a microtube, $t_{\text{b}}$ is the bilayer thickness, $N$ is the number of cylinders inside a tube, and $\langle d \rangle$ is the average distance between walls in a tube.

\Cref{eq:ff_tubes} yields a good fit to the experimental data in the entire $q$-range, except for the intermediate range (\SIrange{0.04}{0.2}{\per\nano\meter}). As shown in FIG. 4A (grey curve), this model captures the oscillations related to the tube radius and the positions of pseudo-Bragg peaks reasonably well. To minimize the discrepancy between the experimental data and the model, an additional term was added to account for the scattering of smaller disk-shaped membrane fragments. The presence of rhombus-shaped nanosheets in cyclodextrin inclusion complex solutions has indeed been previously reported \citep{Yang2017, araujo2022hierarchical} and follows the micelles to vesicle's kinetic model \citep{Guida2010}. To account for their contribution to the sample scattering, the form factor of infinitely thin round disks was added to the form factor of concentric cylinders (Equation \ref{eq:ff_tubes}), which is described by \citet{kratky1949diffuse} as: 
\begin{equation}
    I_\text{disk}(q) \propto \frac{8}{q^2R_d^2}\left(1-\frac{2 J_1(qR_d)}{qR_d}\right),
    \label{eq:ff_disk}
\end{equation}
where $R_d$ is the disk radius. Using the infinitely thin disk form factor is justified as we are interested to model the intermediate $q$ range corresponding to the length scales significantly larger than the bilayer thickness. Introduction of the nanosheet thickness ($\approx t_{\text{b}}$, 3.5 nm) becomes relevant around $q = \frac{2\pi}{t_{\text{b}}} = 1.8$ nm$^{-1}$. However, this value is insignificant for larger length scales, and the simpler expression was used. While the chosen disk shape (round) does not correspond to experimental observations (rhombus), it is the two-dimensional nature of the scattering objects that plays a major role. FIG. 4A (salmon curve) shows that the disk scattering function added to the fitting procedure is constant at low $q$ values and ensures $q$-dependent $-2$ slope at high $q$. 

In general, fitting the model described here to the data allows one to get the quantitative information on the outermost tube radius $\langle R \rangle$, the number of layers inside a tube $\langle N \rangle$, and the distance between them or interbilayer separation $\langle d \rangle$. Gaussian polydispersity up to 10 \% was applied for parameters during the fitting procedure. The size of the flat disks (16-20 nm) used for the modeling did not exceed the microtube radius.

Alternatively, a model allowing for the movement of inner shells within outer shells was suggested to better match the concentric cylinder function with the experimental data. Its best-fit results are presented in the subsection below. However, the model that takes into account other scattering objects contributing at higher $q$ values was found to yield better agreement.

\subsection{Form factor of multiwalled tubes with randomly displaced inner layers}
The concentric cylinders form factor qualitatively describes well key fingerprints from microtube scattering: radius oscillations, interbilayer pseudo-Bragg peaks. However, there is some discrepancy between the model and the experimental points. Scattering intensity is much higher in the intermediate $q$ range near and higher 0.1 nm$^{-1}$. In particular, FIG. 4A shows lamellar peaks that are significantly smeared out on the experimental curve but not on the fit one. The scattering slope in the experiment is less steep than what the model of concentric cylinders predicts. 

Possible explanations of this mismatch include the presence of smaller objects that dominate the sample scattering at intermediate $q$ values. Another reason may be microtube membrane fluctuations. However, the membrane is crystalline,  so we cannot expect many degrees of freedom for it. In addition, microscopy studies on this system have shown a gigantic persistence length \citet{Yang2017}. The third possibility is related to the microtube non-ideality. Namely, cylinders inside a tube can be not precisely concentric. They can shift relative to each other in the cross-section plane. This assumption does not contradict the electron microscopy results presented in the literature \citet{Jiang2010, tang2020supramolecular}. However, using electron microscopy, it is difficult to determine whether such features are inherent in the structure or they appeared during sample preparation for the microscopy.

The following derivation was inspired by the study of \citep{frielinghaus2007small}, where small-angle scattering from multilamellar vesicles with displaced shells was considered. Here we extended this treatment to anisotropic objects - multilayered microtubes. The reported form factor takes into account a possible offset of inner layers inside a tube. The form factor developed here is more computationally expensive compared with the form factor of concentric cylinders.

Assuming that the inner shells can be displaced, let us consider this movement as a random walk. First, we should recall a single cylindrical bilayer shell scattering amplitude:
\begin{widetext}
\begin{equation}
    A\left(R,t_b,q\right) = \frac{RJ_1\left(qR\right)}{q} - \frac{\left(R - t_b\right)J_1\left(q \left(R- t_b\right)\right)}{q}.
    \label{eq:ap9}
\end{equation}

For a multiwalled microtube consisting of $N$ cylindrical shells, we can express a sum of the scattering amplitudes from the shells as follows: 
\begin{equation}
    \sum_{m=0}^{N-1} A_m = \sum_{m=0}^{N-1} A\left(R-md,t_b,q\right)e^{iq\delta x_m}.
    \label{eq:ap10}
\end{equation}

In this expression, $\delta x_m$ is the fluctuation of the position of the $m^{th}$ cylinder relative to the outermost cylinder with radius $R$.

The average intensity for Gaussian random walk:
\begin{equation}\label{eq:ap11}
\left.\begin{aligned} 
\langle \sum_{m=0}^{N-1} A_me^{imq\delta x} \times \sum_{m'=0}^{N-1} A_m'e^{-im'q\delta x}  \rangle &= \sum_{m=0}^{N-1} A_m \times \sum_{m'=0}^{N-1} A_m'\langle e^{-i(m'-m)q\delta x}\rangle   \\
  &  = \sum_{m=0}^{N-1} A_m \times \sum_{m'=0}^{N-1} A_m' K^{|m'-m|},
\end{aligned}\right.
\end{equation}

where $K = e^{\frac{-\sigma ^ 2}{2}}, \sigma = \langle ( q \delta x)^2 \rangle$. 
\end{widetext}

\begin{figure}[tbp]
    \centering
    \includegraphics[width=0.6\textwidth]{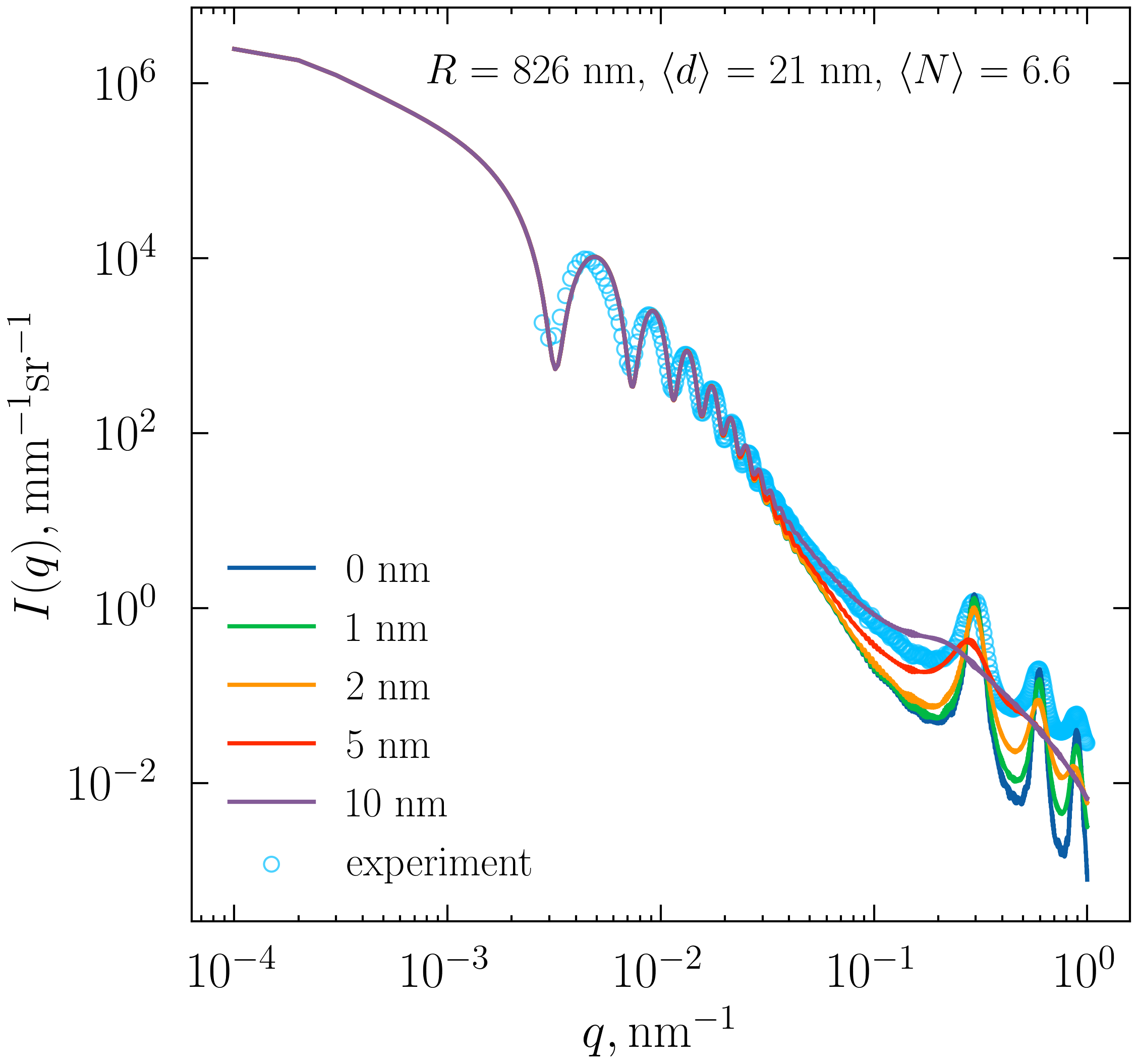}
    \caption{\textbf{Experimental scattering curve ($T$-scan, 10wt\%, 34$^{\circ}$C) (cyan circles) and the modelled form factor with various allowed shifts.} The parameters in the upper right part of the figure were taken from the curve fit by the concentric cylinders form factor. They include outermost radius ($\langle R\rangle $), interbilayer separation ($\langle d\rangle $) values and an average number of cylinders per tube ($\langle N\rangle $). These values were used to model scattering curves with various allowed shift values.}
    \label{fig:shift}
\end{figure}

The resulting expression for the form factor of randomly displaced cylinders was used to model experimental data. \ref{fig:shift} shows experimental scattering data and simulated curves with different average displacements. Because of the sterical reasons, the largest shift cannot exceed $\frac{R}{N}$. When the average shift increases, lamellar peaks of the higher orders are rapidly smeared out. The larger the displacement, the better the model at lower $q$. However, under these conditions, the interbilayer separation is not as clearly defined as for the experimental curve (\ref{fig:shift}, shift 10 nm).
There is something contributing to the electron density fluctuations at the length scales larger than the allowed shift but smaller than the microtube radius. A simulated curve that would be nicely fit the data without a significant scattering mismatch was not found in this simulation procedure. Therefore, this form factor model is believed to be unreasonable for our system.

\subsection{Scattering length densities for the modelling}
X-ray scattering length density (SLD)of a molecule \ce{X} $\rho_{\ce{X}}$ can be seen as a combination of electrons and can be computed using following expression:
\begin{equation}
    \rho_{\ce{X}} = \frac{r_e \times \sum n_i Z_i}{V_m}.
    \label{eq:sld_1}
\end{equation}
Here $r_{\textrm{e}}$ is the Thomson scattering length ($2.818\times10^{-6} \textrm{nm}$), $n_{\textrm{i}}$ is the number density of atom $i$, $Z_{\textrm{i}}$ is the atomic number of atom $i$ (number of electrons), $V_{\textrm{m}}$ is the molecular volume. Basically, the numerator in \cref{eq:sld_1} is a total number of electrons in a molecule multiplied by $r_e$. Regarding SDS@2$\beta$-CD microtubes, we are interested to compute the scattering length density of a bilayer cylindrical shell using \cref{eq:sld_1}. 
To calculate the SLD of 2 $\beta$-cyclodextrin molecules with an embedded SDS tail, we, first, compute a total number of electrons in a pocket. 
\begin{equation}
    N_e(\ce{2\beta-CD})= 2 \times \left(42 \times 6 + 35 \times 8 + 70 \times 1 \right) = 1204.
    \label{eq:N_2betacd}
\end{equation}
\begin{equation}
    N_e(\ce{SDS}^-)= 16 \times 2 + 4\times 8 + 12 \times 6 + 25 \times 1  = 161.
    \label{eq:N_sds}
\end{equation}
After we know the number of electrons in the components, we need the volume occupied by them. The volume occupied per $\beta$-CD molecule is: 
\begin{equation}
    V_{\ce{$\beta$-CD}} =  \frac{1134 \textrm{g}/\textrm{mol} \times 10^{21} \textrm{nm/cm}^3}{1.44 \textrm{g}/\textrm{cm}^3 \times 6.022 \times 10^{23} \textrm{mol}^{-1}} = 1.3 \textrm{ nm}^3. 
    \label{eq:vol}
\end{equation}
The SLD of a pocket consisting of two cyclodextrins and SDS's alkyl chain is around:
\begin{equation}
    \rho_{\textrm{pocket}} = \frac{\left(1204+97\right) \times 2.818\times10^{-6} \textrm{nm}}{2 \times 1.3 \textrm{nm}^3} =  14.1 \times 10^{-4} \textrm{ nm}^{-2}. 
    \label{eq:SLD_pocket}
\end{equation}
The obtained value (Equation \ref{eq:SLD_pocket}) is extremely high. The SDS-head (\ce{SO4-}) is known from the literature: $\rho_{\textrm{SO$_4^-$}} \approx 12.6 \times 10^{-4}\textrm{ nm}^{-2}$ \citep{sanchez2016micelle}. In a bilayer, a small distance between two compexes should have the same SLD as water. The electron density is not homogeneous inside a bilayer, therefore, the step-wise bilayer SLD should be used for the fitting procedure \citep{Yang2017}. 

As we are interested in analyzing the sample scattering that occurs at hundreds of nanometers, the electron density contrast between bilayers and solvent can be simplified. To fit the experimental data we used the constant SLD difference independent of the distance. In general, more precise electron density profile of the bilayer can be derived, though it is not required for $q$ values we are working with. 

\subsection{Absence of polyhedral capsids}

\begin{figure*}[tbp]
    \centering
    \includegraphics[width=1\linewidth]{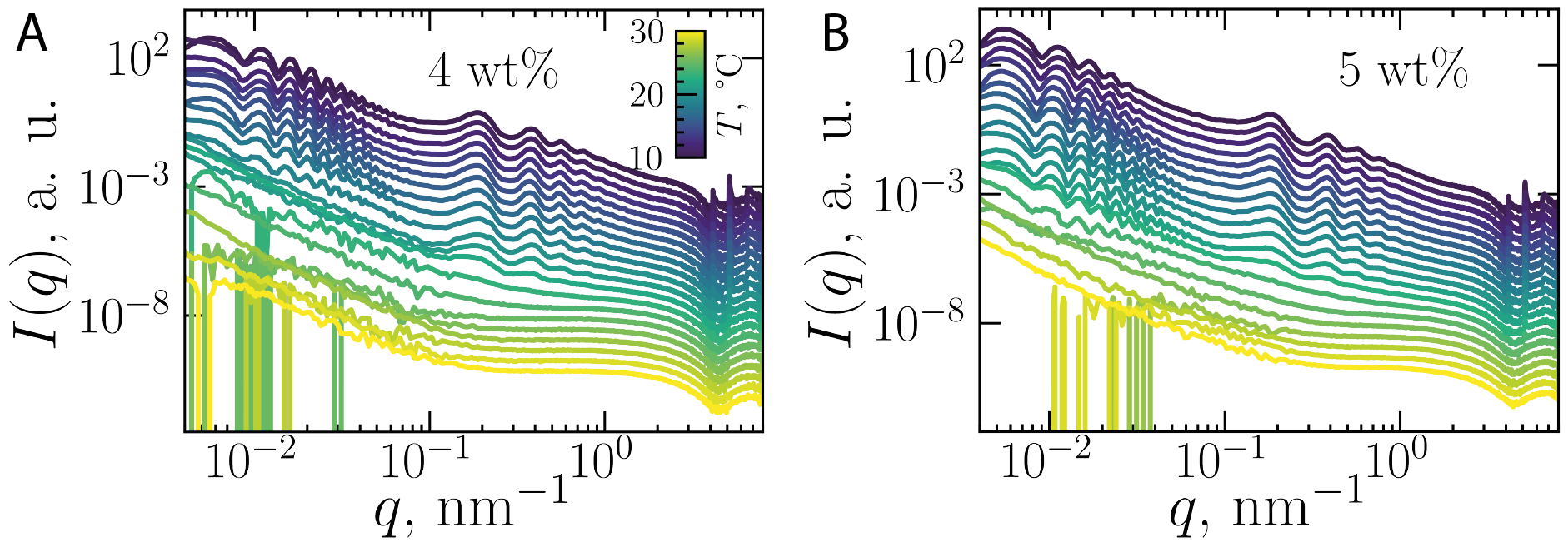}
    \caption{\textbf{SAXS patterns of samples with 4 wt\% and 5 wt\% concentrations recorded during the sample heating.} The curves are offset to track the evolution of scattering fingerprints at different lengthscales. Samples consisting of tubular supramolecular aggregates gradually disintegrate into smaller constituents which is accompanied by the loss of characteristic peaks as well as the loss of scattering intensity at small $q$ values.}
    \label{fig:4_5_wt}
\end{figure*}

\Citet{Jiang2011} has found evidence of a polyhedral capsid phase at $\leq 6$ wt\% in cryo-EM micrographs. We find no evidence for this phase in our SAXS experiments. The proposed fitting model based on concentric cylinders fits well with the experimental data for the samples in the concentration regime $\leq 6$ wt\%, and no capsids were observed. FIG. 4B demonstrates the ability of the proposed model to describe a SAXS curve taken from a 3 wt\% sample at \SI{5}{\celsius}. Moreover, at these low concentrations, scattering curves recorded at different temperatures demonstrate the same fingerprint changes as for samples with higher concentrations, as we show in \ref{fig:4_5_wt}. The melting temperature for this concentration is below room temperature, and no capsids were detected in the SAXS experiment at higher temperatures either --- since the dodecahedral capsids suggested by \citet{Yang2017} possess a crystalline structure, as do the microtubes, we would expect to observe 2D rhombic membrane peaks at high $q$, such as those demonstrated in FIG. 2D. However, these peaks occur only at temperatures below the melting temperature, which we mostly ascribe to microtubes or tiny membrane flakes. Our results suggest that either the capsids appear at negligible concentrations or are artefacts of sample preparation for microscopy characterization.

\section{Polarized light microscopy}

\begin{figure}[tbp]
\centering
\includegraphics[width=1\linewidth]{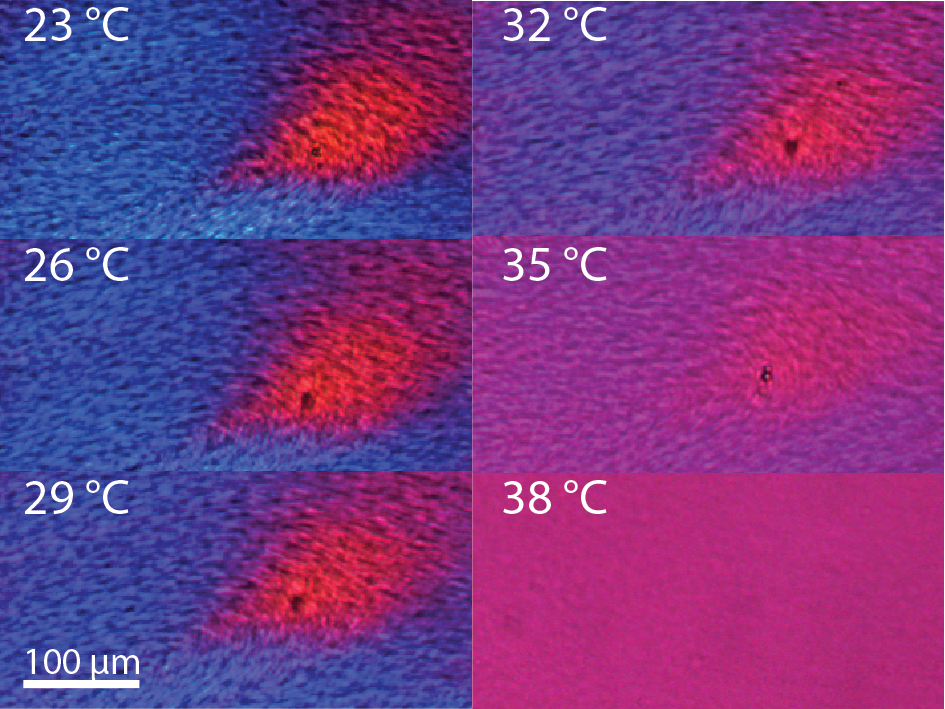}
\caption{\textbf{Gradual melting process of a \qty{10}{wt\percent} sample by PLM.} Frames demonstrate a boundary between two differently oriented domains. The frames were taken as the sample was equilibrated at various temperatures in the range from room $T$ up to $T_{\text{melt}}$. As temperature approaches the melting point, the contrast between domains becomes less pronounced. The sample is immobile, the domains are still present and we assume that they do not change their orientation during the heating process. We attribute the observed changes to the loss of the refractive index gradient because of disassembly of inner layers inside a tube.\label{fig:plm}}
\end{figure}

Polarized light microscopy technique (PLM) can be used for studying samples consisting birefrigent objects. The melting temperature of the birefringent phase and its melting behaviour can be detected. This goal does not require as high resolution as the SAXS technique provides. Since \ce{SDS}@2$\beta$-CD microtubes are anisotropic birefringent objects, we can study their temperature-triggered behaviour using PLM accompanied by the heating stage. A $1/4 \lambda$ retardation plate was used during the experiment to obtain bright, colorful images. 
\ref{fig:plm} displays a boundary between two differently oriented domains, that are, therefore, colored differently. As the temperature increases, the refractive index gradient ($\Delta n$) becomes smaller. Eventually, at high temperatures colors of two domains become the same - birefringence disappears, anisotropic tubes are fully melted. The sample is immobile, the domains are still present and we assume that they do not change their orientation during the heating process. We attribute the gradual changes in $\Delta n$ to the loss of the refractive index gradient because of disassembly of inner layers inside a tube. 

This qualitative observation agrees with the study of \citep{Landman2018InwardMembranes}, where the inward growth mechanism of the SDS@2$\beta$-CD tubes was developed. However, to get quantitative information on the microtube structure, methods with much higher resolution are needed.  

\section{Porod invariant to determine transition point}

\begin{figure}[tbp]
    \centering
    \includegraphics[width=0.6\linewidth]{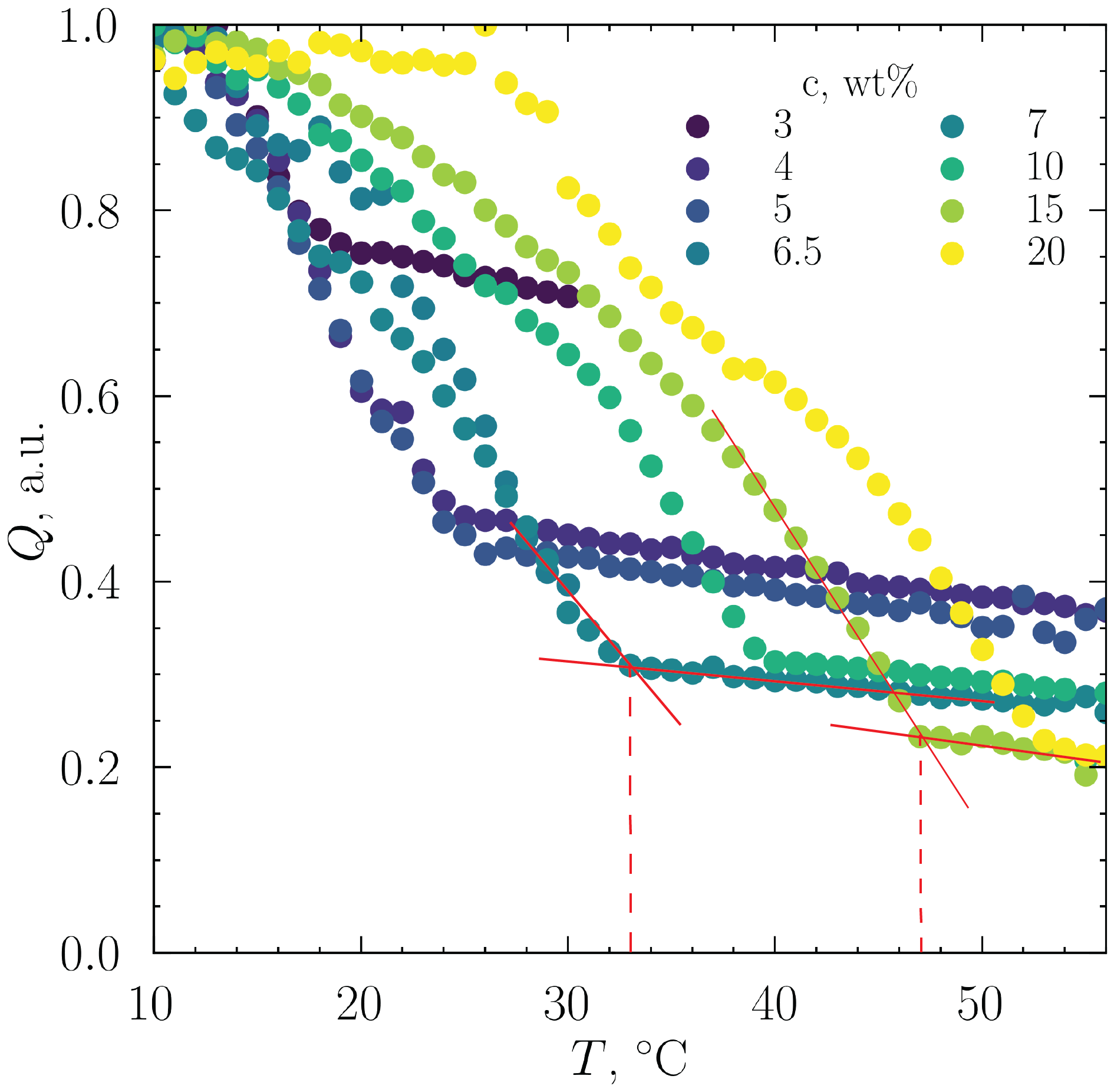}
    \caption{\textbf{Porod invariant value computed from SAXS curves for a set of concentrations and temperatures in the accessible q-region.} The Porod invariant was computed for 0.003 nm$^{-1}$ to 1.7 nm$^{-1}$ $q$ range. Intersection of the red lines were used to determine transition temperature values.}
    \label{fig:Porod}
\end{figure}

Porod invariant, $Q$, is the quantity derived from the Parseval theorem for Fourier transformations. It is computed from the scattering intensity using \cref{eq:porod_gen}. $Q$ provides the mean square electron density contrast. At different temperatures the scattering objects are kept intact, therefore, the mean electron density contrast in the entire $q$ range does not change.
\begin{equation}
    Q = \int_{0}^{\infty} I(q)q^2dq = 2\pi^2 \int_{V} \Delta \rho^2 (\textbf{r}) d( \textbf{r}).
    \label{eq:porod_gen}
\end{equation}
The integral on the left side of \cref{eq:porod_gen} taken from zero to infinity ($q$ values) should always be constant. In a real SAXS experiment, the available $q$ range cannot cover this theoretical limit. But we are able to monitor changes in the mean electron density contrast in a specific scattering vector range $q_{\textrm{min}}$ and $q_{\textrm{max}}$ recorded during the SAXS experiment, where the scattering is governed by the known type of scatterers (\cref{eq:porod}). Fluctuations, detected in the Porod invariant computed in the $q_{\textrm{min}}-q_{\textrm{max}}$ range, refer to the change in the number/size of objects scattering in the considered $q$ range. The scatterers might either agglomerate (then the scattering intensity will contribute to lower $q$) or disintegrate to smaller constituents that scatter at higher $q$.

\begin{equation}
    Q = \int_{q_{\textrm{min}}}^{q_{\textrm{max}}} I(q)q^2dq.
    \label{eq:porod}
\end{equation}

In the case of the microtube melting, we have chosen to consider the lowest available $q$ values, where dominant scattering is supposed to come from large, tubular assemblies. Since the Porod invariant is a measure of the amount of the scattering material, it should decrease as the tubes collapse and disappear (accompanied by the increase in the scattering intensity at higher $q$ values, coming from smaller objects).

\ref{fig:Porod} shows the temperature dependence of computed Porod invariant with a downward trend that holds for all concentrations. $Q$ was computed using \cref{eq:porod} from 0.003 nm$^{-1}$ to 1.7 nm$^{-1}$). In this $q$ region, the scattering features corresponding to the tubular phase for samples of all concentrations are clearly seen. At such $q$ values, scattering is mostly governed by large scattering objects - microtubes. Therefore, a temperature-dependent decrease in the Porod invariant should be associated with the disappearance of these structures. Melting of inner microtube shells starts from lower temperatures, so the Porod invariant becomes smaller. When all microtubes are melted, $Q$ takes the constant minimum value, as there are no more scatters in this $q$ region. Using the intersection between the sharply decaying curve and the flatter line at the higher temperatures (red lines), it is possible to determine the transition temperature of the sample.